\documentstyle[12pt]{article}

\def\beqra{\begin{eqnarray}}
\def\eeqra{\end{eqnarray}}
\def\beqast{\begin{eqnarray*}}
\def\eeqast{\end{eqnarray*}}
\def\be{\begin{enumerate}}
\def\ee{\end{enumerate}}
\def\lag{\langle}
\def\rag{\rangle}

\def\sppt{Research supported in part by the Robert A. Welch
 Foundation and NSF Grant PHY 9009850}

\def\fnote#1#2{\begingroup\def\thefootnote{#1}\footnote{#2}
\addtocounter{footnote}{-1}\endgroup}

\def\beq{\begin{equation}}
\def\eeq{\end{equation}}
\def\haf{\frac{1}{2}}

\def\rf#1{$^{#1}$ }

\begin{document}

\today

\hfill{UTTG-22-92}

\hfill{LBL 33016}

\hfill{UCB 92/36}

\vspace{24pt}

\begin{center}
{\large{\bf   Flavor Changing Scalar Interactions}}

\vspace{36pt}
 Lawrence Hall\fnote{*}{Research supported in part by
NSF grant PHY90-21139 and DOE contract DE-AC03-76SF 00098.}

\vspace{4pt}
Department of Physics, University of California\\
Berkeley, CA, 94720

\vspace{18pt}
Steven Weinberg\fnote{**}{\sppt.}

\vspace{4pt}
Department of Physics, University of Texas\\
Austin, TX, 78712

\vspace{30pt}
{\bf Abstract}

\end{center}

\begin{minipage}{4.75in}

The smallness of fermion masses and mixing angles has
 recently been been attributed to  approximate
global $U(1)$ symmetries, one for each fermion type.
The parameters associated with these symmetry breakings are
 estimated here directly from observed masses and
mixing angles.  It turns out that although flavor changing
 reaction rates may be acceptably small in
electroweak theories
with several scalar doublets without imposing any special
 symmetries on the scalars themselves, such
theories generically yield too much CP violation in the
 neutral
kaon mass matrix.
Hence in these theories CP must also be a good approximate
symmetry. Such models provide an alternative mechanism for CP violation
and have various interesting phenomenological features.
\end{minipage}

\vfill

\baselineskip=24pt
\pagebreak
\setcounter{page}{1}

The inclusion of multiple scalar doublets at the weak scale
in the standard
 $SU(2)\otimes U(1)$ electroweak theory entails the
risk of flavor-changing neutral current processes with rates
 in excess of experimental bounds.  To avoid this, most
studies of such models have adopted the proposal\rf{1}of a
global symmetry that allows only one scalar doublet to
couple to
 the right-handed quarks of each charge.  Recently the need
 for such a symmetry has been challenged in
an article\rf{2}that attributes the various small ratios
 among quark
mixing angles and quark and lepton masses to a set of
 approximate global $U(1)$ symmetries that act only on
the quarks, but not on the scalars.\rf{3}  Specifically, it is
assumed that every appearance of a fermion of the $i$'th
generation in a Yukawa interaction of quarks or leptons with
any
scalar doublet $\phi_n$ is accompanied with a small
dimensionless factor: $\epsilon_{Q_i}$ for left-handed
quark doublets; $\epsilon_{U_i}$ or $\epsilon_{D_i}$ for
right handed quarks; $\epsilon_{L_i}$ for left-handed
lepton doublets; and $\epsilon_{E_i}$ for right-handed
charged leptons.  That is, writing the general
Yukawa interaction in the form
\beq
{\cal L}_Y=-\lambda_{ijn}^U \bar{Q}_{Li} U_{Rj}\cdot
\tilde{\phi}_n-\lambda_{ijn}^D \bar{Q}_{Li} D_{Rj}
\cdot \phi_n-\lambda_{ijn}^E \bar{L}_{Li} E_{Rj}
\cdot\phi_n+H.c.,
\eeq
$$
Q_{Lj}\equiv\left[\begin{array}{l}U_{Lj}\\D_{Lj}\end{array}
\right]\quad \phi_n\equiv \left[\begin{array}{l}\phi_n^
+\\ \phi_n^0\end{array}\right]
\quad
\tilde{\phi}_n\equiv
\left[\begin{array}{r}\phi_n^{0*}\\-\phi_n^{+*}
\end{array}\right]
$$
the Yukawa couplings are assumed to be of order
\beq
|\lambda_{ijn}^U| \approx \epsilon_{Q_i}\epsilon_{U_j}\qquad
|\lambda_{ijn}^D| \approx \epsilon_{Q_i}\epsilon_{D_j}\qquad
|\lambda_{ijn}^E| \approx \epsilon_{L_i}\epsilon_{E_j}
\eeq
for all $n$.
(Here and below, "$\approx$" will be understood to indicate
equality within a factor of order two or three.)
Though there is no compelling theoretical justification
for this assumption, it may be taken as representative
of any theory of fermion-scalar couplings that
attributes the small fermion masses and mixing angles
to symmetries that act on the fermions rather than
the scalars.  With the aid of an additional  somewhat
  ad hoc ansatz relating the
$\epsilon$'s, it was shown in reference 2 that the rates of
flavor-changing neutral current processes can be kept
within experimental bounds without invoking any symmetry
that restricts which scalars can interact with which quarks.
We shall recover the same result here without using this
ansatz.
But as we shall see, there is a further problem with such
multi-scalar models: they do not necessarily yield small
violations of $CP$-conservation in the neutral kaon system.

To analyze this problem, the generations
will be ordered so that, for $i<j$,
\beq
\epsilon_{Q_i}\leq \epsilon_{Q_j}\qquad
\epsilon_{U_i}\leq \epsilon_{U_j}\qquad\epsilon_{D_i}\leq
\epsilon_{D_j}\qquad\epsilon_{L_i}\leq \epsilon_{L_j}\qquad
\epsilon_{E_i}\leq \epsilon_{E_j}.
\eeq
The mass matrices arising from (1) may then be put into a
real diagonal form by subjecting the fermions
to transformations:
\beqra
&& U_{Li}\rightarrow V^{U_L}_{ij} U_{Lj}
\qquad  D_{Li}\rightarrow V^{D_L}_{ij}D_{Lj}\nonumber\\&&
U_{Ri}\rightarrow V^{U_R}_{ij}U_{Rj}\qquad
D_{Ri}\rightarrow V^{D_R}_{ij}D_{Rj}\nonumber\\&&
E_{Li}\rightarrow V^{E_L}_{ij}E_{Lj}\qquad E_{Ri}\rightarrow
V^{E_R}_{ij}E_{Rj},
\eeqra
with unitary matrices $V^{U_L}_{ij}$, etc., having elements
\beq
V^{U_L}_{ij}\approx\left\{\begin{array}{ll}
\epsilon_{Q_i}/\epsilon_{Q_j}\qquad & i\leq j\\
\epsilon_{Q_j}/\epsilon_{Q_i}\qquad & j\leq
i\end{array}\right.,
\eeq
and likewise for $V^{D_L}_{ij}$, $V^{U_R}_{ij}$,
$V^{D_R}_{ij}$, $V^{E_L}_{ij}$, and $V^{E_R}_{ij}$.
This transformation yields quark and lepton masses of order
\beq
m_{U_i}\approx
\epsilon_{Q_i}\epsilon_{U_i}\lag\phi\rag\qquad
m_{D_i}\approx
\epsilon_{Q_i}\epsilon_{D_i}\lag\phi\rag\qquad
m_{E_i}\approx \epsilon_{L_i}\epsilon_{E_i}\lag\phi\rag
\eeq
and a Cabibbo-Kobayashi-Maskawa (CKM) matrix of the form
\beq
V_{ij}\equiv [V^{U_L\dagger} V^{D_L}]_{ij}\approx
\left\{\begin{array}{ll} \epsilon_{Q_i}/\epsilon_{Q_j}\qquad
& i\leq j\\
\epsilon_{Q_j}/\epsilon_{Q_i}\qquad & j\leq
i\end{array}\right.,
\eeq
where $\lag\phi\rag$ is the common order of magnitude of the
{\em conventionally} normalized complex neutral scalars,
of order 247 GeV/$\sqrt{2} =175$ GeV.

Now we will use experimental data to estimate the
$\epsilon$'s.
First, the ratios of the $\epsilon_{Q_i}$ are directly given
by Eq. (5) in terms of the mixing angles.  The ratio
$\epsilon_{Q_1}/\epsilon_{Q_2}$ may be determined either
from the Cabibbo angle
$$ \epsilon_{Q_1}/\epsilon_{Q_2}\approx V_{us}= 0.218\; {\rm
to} \; 0.224$$
or less accurately  from semi-leptonic $B$ meson
decays\rf{4}
$$  \epsilon_{Q_1}/\epsilon_{Q_2}\approx {V_{ub} \over
V_{cb}} \simeq 0.07 \;.$$
Given the theoretical uncertainties in extracting the ratio
$V_{ub}/ V_{cb}$, we regard these two estimates as being
satisfactorily consistent, and we take
$\epsilon_{Q_1}/\epsilon_{Q_2} = 0.2$. The second ratio of
$\epsilon_{Q_i}$ is
determined from
 $$ \epsilon_{Q_2}/\epsilon_{Q_3}\approx V_{cb}= 0.032\;
{\rm to}\; 0.054\;.$$
Hence we take
\beq
\epsilon_{Q_1}/\epsilon_{Q_2}\approx .2\qquad
\epsilon_{Q_2}/\epsilon_{Q_3}\approx .04
\qquad \epsilon_{Q_1}/\epsilon_{Q_3}\approx .008\;.
\eeq
Using (8), (6), and the ``experimental'' values of the
quark masses\rf{5}, we have then also
\beqra
&&\epsilon_{U_1}\approx .004 / \epsilon_{Q_3}
\qquad\epsilon_{U_2}\approx .2/ \epsilon_{Q_3}\\
&&\epsilon_{D_1}\approx .006 /\epsilon_{Q_3}\qquad
\epsilon_{D_2}\approx .025 /\epsilon_{Q_3}
\qquad \epsilon_{D_3}\approx .03
/\epsilon_{Q_3}\;.\nonumber\\&&{}
\eeqra
The Yukawa couplings in Eq. (1) can then be estimated from
Eq. (2), with the unknown $\epsilon_{Q_3}$ cancelling in all
couplings.

Though it is not needed in estimating the Yukawa couplings,
we can also estimate the factor $\epsilon_{Q_3}$
which is needed to determine the individual suppression
factors.  The top quark mass cannot be much less than
$\lag\phi\rag\simeq $ 175 GeV, so if either of the
quantities $\epsilon_{Q_3}$ and $\epsilon_{U_3}$ were
much smaller than the other, then the larger would have to
be much larger than unity, contrary to our assumption that
the $\epsilon$'s are {\em suppression} factors.
Thus Eq. (6) indicates that
$\epsilon_{Q_3}\approx\epsilon_{U_3} \approx
\sqrt{m_t/\lag\phi\rag}$.  But this actually
applies to the Yukawa couplings defined at a renormalization
scale of $m_t$, while we choose to quote the couplings
defined at a renormalization scale of 1 GeV,
which are larger by a factor $Z\approx 2$.  We
therefore estimate
\beq
\epsilon_{Q_3}\approx \sqrt{Z m_t/\lag\phi\rag}
\eeq
it being understood from now on that all $\epsilon$'s  are
defined at a renormalization scale of order 1 GeV.

With no measurable mixing angles for leptons, we cannot
determine separate values for the leptonic
suppression factors $\epsilon_{E_i}$ and $\epsilon_{L_i}$.
However the most stringent limits on scalar interactions
were found
in reference 2 to be set by the non-leptonic
$K^0\leftrightarrow \bar{K}^0$ and $B^0\leftrightarrow
\bar{B}^0$ transitions, to
which we now turn.  (The transitions $D^0\leftrightarrow
\bar{D}^0$ and  $B_s^0\leftrightarrow \bar{B_s}^0$
will be considered later.)

Exchange of neutral scalars produces two different
kinds of
parity-conserving $\Delta S=2$ four-quark operators that can
induce the transition $K^0\leftrightarrow \bar{K}^0$:
\beq
{\cal L}_{\Delta S=2}=2G(\bar{s}_Rd_L)(\bar{s}_Ld_R)+G'
\left[(\bar{s}_Ld_R)^2+(\bar{s}_Rd_L)^2\right]
\eeq
with coupling constants
\beq
G=\sum_{nmN}\lambda^{D*}_{12n}
\lambda^D_{21m}A_{nN}A^*_{mN}/m_N^2
\eeq
\beq
 G'=\haf\sum_{nmN}[\lambda^D_{21n}
\lambda^D_{21m}A_{nN}A_{mN}+\lambda^{D*}_{12n}\lambda^{D*}
_{12m}A_{nN}^*A_{mN}^*]/m_N^2
\eeq
where
\beq
\lag 0| \phi^0_n(0) |N\rag\equiv
\frac{A_{nN}}{(2\pi)^{3/2}\sqrt{2\omega_N}}
\eeq
and the sum over $N$ runs over neutral Higgs scalar mass
eigenstates.  For an order-of-magnitude estimate of
the $K_1^0\;-\;K_2^0$ mass difference produced by this
interaction, we will fall back on the vacuum
insertion approximation (which is justified in quantum
chromodynamics in the limit of a large number of colors):
\beq
\lag \bar{K}^0|{\cal O}_1 {\cal O}_2|K^0  \rag \approx
\lag \bar{K}^0|{\cal O}_1|0\rag \lag 0| {\cal O}_2| K^0 \rag
+\lag \bar{K}^0|{\cal O}_2|0\rag \lag 0| {\cal O}_1| K^0 \rag
\eeq
where each of ${\cal O}_1$ and ${\cal O}_2$ is either
$(\bar{s}_Ld_R)$ or $(\bar{s}_Rd_L)$.  The
one-particle matrix elements of these operators can be
calculated
from the known matrix elements of the corresponding
axial-vector current:
\beq
\lag 0| (\bar{s}_Rd_L)| K^0\rag = -\lag 0| (\bar{s}_Ld_R)|
K^0\rag = \frac{m_K^2 F_K}{(2\pi)^{3/2}\sqrt{2m_K}2\sqrt{2}m_s}
\eeq
where $F_K\simeq 230$ MeV is the kaon decay amplitude (as
compared with $F_\pi\simeq 190$ MeV.)
This gives a $K_1^0\:-\:K_2^0$ mass difference
\beq
\Delta M_K\approx \frac{(G-G')m_K^3F_K^2}{4m_s^2}\;.
\eeq
The flavor-changing suppression factors in $G$ and $G'$
turn out to be about the same
\beq
\epsilon_{Q_1}\epsilon_{D_2}\epsilon_{Q_2}\epsilon_{D_1}
\approx
\haf[\epsilon^2_{Q_2}\epsilon^2_{D_1}+\epsilon^2_{Q_1}
\epsilon^2_{D_2}]\approx 5\times 10^{-8}\;.
\eeq
The $A_{nN}$ are of order unity, so barring unexpected
cancellations, we expect that
\beq
G-G'\approx  5\times 10^{-8}\:e^{i\delta}/m_H^2
\eeq
where $\delta$ is an unknown phase, and $m_H$ is a weighted
average of neutral scalar masses.  Using this in (18)
[with $m_s\simeq 180$ MeV] then yields
\beq
|\Delta M_K|\approx \frac{5\times 10^{-
8}m_K^3F_K^2}{4m_s^2m_H^2}\approx \frac{3\times 10^{-5}\:
{\rm
eV}}{(m_H/300\, {\rm GeV})^2}\;.
\eeq

The analysis we use to estimate $\Delta M_{B}$ parallels
that used in Eqs. (12) to (21) for
$\Delta M_K$.  The relevant coupling suppression factors are
now
 \beq
\begin{array}{cc}
(\bar{b}_Ld_R)(\bar{b}_Rd_L)
& \qquad \epsilon_{D_3}\epsilon_{Q_1}\epsilon_{D_1}
\epsilon_{Q_3}\approx  10^{-6}\\{} & {} \\
\haf[(\bar{b}_Ld_R)^2+(\bar{b}_Rd_L)^2] & \qquad
\haf[\epsilon^2_{D_3}\epsilon^2_{Q_1}+
\epsilon^2_{D_1}\epsilon^2_{Q_3}]\approx 2\times 10^{-5}
\end{array}
\eeq
(Note that the second suppression factor is an order of
magnitude larger than the naive estimate $m_dm_b/\lag \phi
\rag^2$.)   There have been many theoretical estimates of
$F_B$,  summarized by Buras and Harlander\rf{6}.  As a rough
consensus value, we shall take $F_B\approx 230 $ MeV.
Following the same reasoning as for $\Delta M_K$, we have
then
\beq
|\Delta M_B|\approx \frac{2\times 10^{-
5}m_B^3F_B^2}{4m_b^2m_H^2}\approx \frac{ 10^{-2}\:
{\rm
eV}}{(m_H/300\, {\rm GeV})^2}\;.
\eeq

There are also the more familiar box diagrams involving $WW$
exchange.  Assuming no accidental
cancellations between these contributions, it seems
reasonable
to
require that the scalar exchange contributions should
not exceed twice the experimental values, $|\Delta M_K|=3.5
\times 10^{-6}$ eV and
$|\Delta M_{B}| = (3.6 \pm 0.7) \times 10^{-4} $ eV.  This
yields the conditions $m_H>600 $ GeV and
$m_H>1 $ TeV, for $K$ and $B$, respectively.

These Higgs masses are somewhat larger than seems
plausible, but our analysis
involves a number of dubious approximations, and it
is entirely possible that we have overestimated the
matrix elements for $K^0\leftrightarrow \bar{K}^0$
and $B^0\leftrightarrow
\bar{B}^0$ transitions by factors of two or three.
We conclude then  that, as found in reference 2, the selection
rule of reference 1 is not indispensable in
keeping the scalar exchange contribution to the
$K^0_1\:-\:K^0_2$ and $B_1^0\:-\:B_2^0$ mass
differences at a reasonable size.  But without this selection
rule the Higgs scalars must be relatively heavy, and even
so scalar exchange would be likely to make a large and
perhaps dominant contribution to the $K^0_1\:-\:K^0_2$ and $B_1^0\:-\:B_2^0$
mas
s
differences.

Our conclusions shift when we consider the CP-violating part of the
$K^0_1\:-\:K^0_2$ mass difference.  This is usually
expressed in
terms of a parameter $\epsilon$ with $|\epsilon| \simeq
2.26\times 10^{-3}$, which (for
$|\epsilon'|\ll |\epsilon|)$ is given by
$\epsilon={\rm Im}(\Delta M_K)/\sqrt{2}|\Delta M_K|$.  If
scalar
exchange does indeed make a major
contribution to the $K^0_1\:-\:K^0_2$ mass difference, then
the
phase $\delta$ in Eq. (20) would have to be quite small, of
the order of
a milliradian or less, in contradiction with the
general expectation that all phases are of order unity.
This leaves us with an interesting choice of alternatives:
\begin{itemize}
\item The CP-violating phases are indeed generically of
order unity, but scalar exchange contributions to the
$K^0_1\:-\:K^0_2$ mass difference are much smaller than we
have estimated, perhaps because of accidental cancellations
in the calculation of the scalar-exchange contribution
to the four-quark operator, or a gross failure of the vacuum
insertion approximation used in calculating the
$K^0\;-\;\bar{K}^0$ matrix element, or both.  This seems
implausible unless the scalars are very heavy.
\item
The CP-violating phases are generically of order unity,
but the scalar couplings {\em are} constrained by
the selection rule of reference 1.  This is of course
automatic with just one scalar doublet, or in supersymmetric
theories with just two scalar doublets.  (However it is not
at all automatic in supersymmetric theories
with more than two scalar doublets.  In particular, if
several scalar doublets couple to the right-handed quarks of
charge $-1/3$, and if as usually assumed these
scalars have smaller
vacuum expectation values than the doublets that couple to
the right-handed quarks of charge $2/3$, then the
Yukawa couplings of these scalars would be correspondingly
larger, leading to an even larger
$K^0_1\:-\:K^0_2$ mass difference.)
\item All of the estimates in this paper are valid, but
CP is a good approximate symmetry, with all the CP-
violating phases like $\delta$ of order $10^{-3}$.
\end{itemize}

The third alternative is admittedly a somewhat reactionary
view of CP nonconservation.  After the discovery of the
process $K^0_2\rightarrow\pi+\pi$       in 1964 it was
widely assumed that this process is much slower than
$K^0_1\rightarrow\pi+\pi$ because CP is a good approximate
symmetry for the weak interactions.  Then following the
discovery of a third generation of quarks and leptons in the
1970s, physicists became attracted to the idea that
CP-violating phases are typically of order unity, and that
CP only seems to be a good approximate symmetry because the
third generation is  weakly mixed with the first two.
However,  since we know that in any case we have to deal
with quark masses and mixing angles that for mysterious
reasons are very small, there is nothing absurd in supposing
that CP-violating angles are also small.  Indeed, apart from
any consideration of scalar exchange effects, we may be
driven to this assumption if  theories with
supersymmetry broken at the electroweak scale prove
successful. Such theories have CP violating
phases in the supersymmetry breaking interactions that
generically lead to a
neutron electric dipole moment three orders of magnitude
larger than the
present experimental limit\rf{7}\nolinebreak. This major
problem of supersymmetrics
models is avoided if we assume that CP-violating phases are
generically of  order $10^{-3}$. In the balance of this
paper we discuss the
experimental consequences of this picture of CP violation,
combined (where relevant) with our earlier assumptions
regarding scalar couplings.

(1) Direct CP violating effects in the decays of $K$ mesons
will be unobservably
small. The CKM contribution to $|\epsilon'/\epsilon|$ will
be of order
$10^{-6}$,
and the contribution from tree level exchange of scalar
mesons will be even
smaller. Hence these theories predict that the next round of
experiments at
CERN and Fermilab will {\em not}  find a signal for
$|\epsilon'/\epsilon|$ at the
projected level of sensitivity of $10^{-4}$. Such a null
result would be
extremely exciting since it would imply that the CKM matrix
could not be the
origin of the known CP violation (unless the top quark mass
is found to take
a value allowing a precise cancellation between two
contributions to
$\epsilon'/\epsilon$), thus implying an alternative source
of CP violation, such as scalar exchange.

(2) All CP violating asymmetries which arise in particle
decays must be of order
$10^{-3}$ or less, since these asymmetries must be
proportional to a CP
violating phase. In particular
CP violating effects in $B$ meson decays will be too small
to be observed in any
experiment proposed to date. For example the angles $\alpha,
\beta$ and
$\gamma$ of the unitarity triangle of the CKM matrix will be
of order $10^{-3}$
and will be far too small to be observed at proposed $B$
factories.
Nevertheless such $B$ factories could definitively exclude
the CKM origin of CP
violation\rf{8}.

(3) The most promising new positive signature of CP
violation in our scheme is the
neutron electric dipole moment. The electric dipole moment
of the up quark
arises from a one loop diagram with a virtual top quark and
Higgs meson, and
using the results of eqs. 8, 9 and 10 we estimate  the
resulting neutron
electric dipole moment to be of order $10^{-26}\;e$ cm,
close to the current
experimental limit. In supersymmetric theories a comparable
contribution would
be expected from diagrams with internal superpartners. The
electron electric
dipole moment is expected to be of order $10^{-31}\; e$ cm.

(4) The predictions for the branching ratios for many rare K
meson decays are not
the same in our scheme as in the standard model. The most
drastic change is for
the $K^0_2 \rightarrow \pi \nu \bar \nu$ amplitude which is
proportional to
the CKM CP violating phase and therefore gets suppressed by
two to three orders
of magnitude. There is no tree level Higgs exchange
contribution to this decay
because the Higgs mesons do not couple to neutrinos.

(5) It is
striking that for Higgs bosons with a typical mass of about
700 GeV and with
couplings to quarks determined by Eqs. (8), (9) and (10),
the tree level scalar exchange contribution
to neutral $K$ and
$B$ meson mass mixing turned out to be at about the level
observed by experiment.  Although this
means that little can be learned about the CKM matrix from
$\Delta M_K$ and
$\Delta M_{B}$, the case of $D$ - $\bar{D}$ presents
different opportunities.  The analysis we use to estimate
$\Delta M_D$ parallels that used in Eqs. (12) to (21) for
$\Delta M_K$ and $\Delta M_B$.  The relevant coupling
suppression factors are now
 \beq
\begin{array}{cc}
(\bar{c}_Lu_R)(\bar{c}_Ru_L) & \qquad
\epsilon_{U_1}\epsilon_{Q_2}\epsilon_{U_2}
\epsilon_{Q_1}\approx 3\times 10^{-7}\\{} & {} \\
\haf[(\bar{c}_Lu_R)^2+(\bar{c}_Ru_L)^2] & \qquad
\haf[\epsilon^2_{U_1}\epsilon^2_{Q_2}+
\epsilon^2_{U_2}
\epsilon^2_{Q_1}]\approx 1\times 10^{-6}\\{} & {} \\
\end{array}
\eeq
  A theoretical estimate of $F_D$ may be obtained from
the previously quoted estimate $F_B \simeq 230 $ MeV,
using the relation (valid in the limit of large
quark masses) $F_D/F_B\simeq \sqrt{m_b/m_c}$.  This
gives $F_D\approx 470$ MeV, so that
\beq
|\Delta M_D|\approx \frac{ 10^{-
6}m_D^3F_D^2}{4m_c^2m_H^2}\approx \frac{2\times 10^{-3}\:
{\rm
eV}}{(m_H/300\, {\rm GeV})^2}\;.
\eeq
If we take the typical Higgs mass as near 1 TeV to account for
the observed values of $|\Delta M_K|$ and $|\Delta M_B|$,
then the predicted value of $|\Delta M_D|$ is close to the
current experimental limit, $|\Delta M_D|< 1.3\times 10^{-
4}$ eV.    In
the standard model $\Delta M_D$ is dominated by long
distance contributions,
which were originally estimated\rf{9} to be in the range
(0.3 to  0.01)$\times
10^{-4}$ eV, very much larger than the order $10^{-8}$eV
contribution
from the short distance standard model box diagram.
In this case, a positive
observation of mass mixing at the level of $10^{-4}$ eV
would not necessarily
require new physics beyond the standard model.  However a
recent
study\rf{10}using heavy quark effective field theory and
naive dimensional analysis
suggests that the long distance standard model contribution
to $\Delta M_D$ is in fact
only modestly (about an order of magnitude) larger than the
short distance
contribution. Furthermore, a subsequent calculation\rf{11},
which includes leading order QCD corrections, supports this
low value of $\Delta M_D$ in the standard model.
On this basis,  we can conclude that a
positive signal of neutral $D$
meson mixing at the next round of searches at Fermilab, CESR
and a tau/charm
factory would provide evidence in favor of our scheme.

(6) For strange neutral beauty meson mixing
$B^0_s\leftrightarrow \bar{B}^0_s$ transitions,
 the relevant suppression factors are
 \beq
\begin{array}{cc}
(\bar{b}_Ls_R)(\bar{b}_Rs_L)
& \qquad \epsilon_{D_3}\epsilon_{Q_2}\epsilon_{D_2}
\epsilon_{Q_3}\approx 3\times 10^{-5}\\{} & {} \\
\haf[(\bar{b}_Ls_R)^2+(\bar{b}_Rs_L)^2] & \qquad
\haf[\epsilon^2_{D_3}\epsilon^2_{Q_2}+
\epsilon^2_{D_2}\epsilon^2_{Q_3}]\approx 3\times 10^{-4}\;.
\end{array}
\eeq
 Assuming that the experimental value of $\Delta M_{B}$ is
dominated by scalar exchange, the scalar-mediated
contribution to $B_s$
mixing is predicted to be of order
\beq
\left( \Delta M_{B_s}\right)_{scalar} \approx \left(
{\epsilon^2_{D_3}\epsilon^2_{Q_2}+
\epsilon^2_{D_2}\epsilon^2_{Q_3}} \over
\epsilon^2_{D_3}\epsilon^2_{Q_1}+
\epsilon^2_{D_1}\epsilon^2_{Q_3}\right) \Delta M_{B} \approx
5\times 10^{-3}\,{\rm eV}\;.
\eeq

%\beq
%\begin{array}{cc}
%(\bar{c}_Lu_R)(\bar{c}_Ru_L) & \qquad
%\epsilon_{U_1}\epsilon_{Q_2}\epsilon_{U_2}
%\epsilon_{Q_1}\approx 3\times 10^{-7}\\{} & {} \\
%\haf[(\bar{c}_Lu_R)^2+(\bar{c}_Ru_L)^2] & \qquad
%\haf[\epsilon^2_{U_1}\epsilon^2_{Q_2}+
%\epsilon^2_{U_2}
%\epsilon^2_{Q_1}]\approx 1\times 10^{-6}\\{} & {} \\
%(\bar{b}_Ld_R)(\bar{b}_Rd_L)
%& \qquad \epsilon_{D_3}\epsilon_{Q_1}\epsilon_{D_1}
%\epsilon_{Q_3}\approx 1\times 10^{-6}\\{} & {} \\
%\haf[(\bar{b}_Ld_R)^2+(\bar{b}_Rd_L)^2] & \qquad
%\haf[\epsilon^2_{D_3}\epsilon^2_{Q_1}+
%\epsilon^2_{D_1}\epsilon^2_{Q_3}]\approx 2\times 10^{-5}
%\end{array}
%\eeq

(7) In theories with only one scalar
doublet coupling to quarks of a given charge,\rf{1} the
positively charged scalars decay
predominantly
to $c \bar{s}$ and $\nu_\tau \bar{\tau}$, when the $t \bar{b}$ mode is
kinematically  forbidden.
In the present
class of theories the
decay to $c \bar{b}$ completely dominates because the
relevant products of
$\epsilon_i$ are more than an order of magnitude larger for
this mode than any
other.

(8) Finally we consider exotic decay modes of the top quark.
Our estimates indicate that in the class of theories we are
considering  Higgs particles would be too heavy to appear
among the decay products of top quarks.  But the
phenomenology of such decays would be quite interesting, so
it is worth considering the possibility that we have
seriously overestimated neutral meson mass mixing, and that there are
some Higgs scalars lighter than the top quark.  In most
models with more than a single scalar doublet the exotic
decays
$t \rightarrow b h^+$ and $t \rightarrow c h^0$ will occur
if they are kinematically allowed. (Here $h^+$ and $h^0$
are the lightest non-Goldstone mass eigenstates formed from
linear combinations of the scalars destroyed by the fields
$\phi_n^+$ and $\phi_n^0$ introduced in eq. 1.)  As
indicated above, the $h^+$ would decay predominantly through
the channel $h^+\rightarrow c \bar{b}$, and the $h^0$ decays predominantly via
$h^0\rightarrow b \bar{b}$, so that either of these exotic
top quark decays yields $t \rightarrow b \bar{b} c$.
However, as will be discussed below, the $h^0$ also has a large branching ratio
to tau pairs.

The decays $t \rightarrow b h^+$ are induced by the Yukawa
interaction $\lambda_{33n}^U \bar{Q}_{L3} U_{R3}\cdot
\tilde{\phi}_n$, leading to a decay rate
\beq
\Gamma(t \rightarrow b h^+) \approx {G_F m_t^3 \over 8
\sqrt{2} \pi}
\left( 1 - {m_{h^+}^2 \over m_t^2} \right)^2
\eeq
If 46 GeV $\leq m_t \leq M_W$ then this exotic decay mode
would dominate all others by a
large factor, explaining how a top quark with mass less than
$m_W$ might not have been discovered. The charged
Higgs $h^+$ decays
predominantly to $c \bar{b}$. Using our
values of the $\epsilon_i$ we compute
the branching ratio to $\bar{\tau} \nu_\tau$ to be
only $\approx 10^{-3}$.
Hence, in this class of theories a
successful search for the top quark at the Fermilab collider
would require a
technique to isolate candidate events with four b-type
quarks and up to
six jets.  On the other hand, if $m_t > M_W$ we find
\beq
{\Gamma(t \rightarrow b h^+) \over \Gamma(t \rightarrow b
W^+)} \approx
\left({1 - {m_{h^+}^2 \over m_t^2} \over 1 - {M_W^2 \over
m_t^2}}\right)^2
{1 \over 1 + 2{M_W^2 \over m_t^2}}
\eeq
which implies that a significant suppression of the
conventional decay
mode can occur. For example for a top quark
mass of 100 GeV and a scalar mass of 50 GeV the conventional
isolated lepton
signature of the top quark will be suppressed by a factor of
about 3. With
sufficient statistics the top quark can still be discovered
by the conventional
mode, although a determination of its mass from the rate of
these events could
result in a considerable overestimate, about 25 GeV in the
example given above.
For $m_t \geq$ 150 GeV the suppression of the conventional
signal will be a
factor of two or less.

Turning to the decay $t\rightarrow c
h^0$, we note that this decay
  is
of great interest since, unlike the decay to $b h^+$, this
flavor-changing
decay mode can only be large if the symmetry imposed in reference 1
is relaxed\rf{12}.
This decay is induced by the operator $\lambda_{32n}^U
\bar{Q}_{L3} U_{R2}\cdot
\tilde{\phi}_n$. The relevant coupling factor
$\epsilon_{Q_3}
\epsilon_{U_2} \approx 0.2$ is surprisingly large in this
case\rf{13} and such
decays dominate (aside from the possible decay $t\rightarrow
b h^+$) if the top quark is lighter than the $W$ boson.
The neutral Higgs $h^0$ decays predominantly to $\bar{b} b$. Using our values
for the $\epsilon_i$ we find the branching ratio to tau pairs to be $\approx
10^{-1}$. Thus $h^0$ has much larger leptonic branching ratios than $h^+$. We
expect the best signature at the Fermilab collider to occur when one neutral
Higgs decays to b pairs and the other to tau pairs, with one tau giving an
isolated electron and the other an isolated muon. For an integrated luminosity
of $10 pb^{-1}$ and a top quark mass of 80 GeV, the Fermilab collider would
produce $\approx 30$ such events, with a signature $e + \mu +$ jets
(from 2b and 2c quarks) + missing transverse energy. A search for these events
must take into account the softer $p_T$ distribution of the isolated leptons
compared to the distribution expected from conventional top quark decays.

For the case $m_t > M_W$, the exotic decay mode is no longer likely to dominate
\beq
{\Gamma(t \rightarrow c h^0) \over \Gamma(t \rightarrow b
W^+)} \approx
{\epsilon_{U_2}^2 \over \epsilon_{U_3}^2}
\left({1 - {m_{h^0}^2 \over m_t^2} \over 1 - {M_W^2 \over
m_t^2}}\right)^2
{1 \over 1 + 2{M_W^2 \over m_t^2}}\;.
\eeq
The decay $t \rightarrow c h^0$ does not significantly deplete
the conventional decays, so the discovery of the top
quark is not hindered by this process. However the discovery
of such
exotic, flavor-changing decays  would not only
reveal a Higgs boson but would strongly suggest a theory of
several scalar
doublets with approximate flavor and CP symmetries.

We are grateful for conversations with Howard Georgi.

\pagebreak

\noindent
{\bf References}

\begin{enumerate}
%1
\item
S. L. Glashow and S. Weinberg, Phys. Rev. D15, 1958 (1977).
%2
\item
A. Antaramian, L. J. Hall, and A. Ra\v{s}in, Phys. Rev. Lett.
69 1871 (1992).
%3
\item
The necessity of coupling scalars according to the rules
of reference 1 was also questioned by H. Georgi and D. V.
Nanopoulos, Physics Letters 82B, 95 (1979).  They accounted
for the suppression of flavor changing processes like
$K^0\leftrightarrow \bar{K}^0$ by supposing that the scalars
with flavor changing couplings are much heavier than the
Higgs boson responsible for electroweak symmetry breaking.
Here we assume that all scalars have roughly comparable
masses, and attribute the suppression of flavor changing
processes to the smallness of known quark masses and mixing
angles.
%4
\item
P. Drell, Talk presented at the International Conference on
High Energy
Physics, Dallas. Aug. 1992.
%5
\item
We use $m_u=5.5$ MeV, $m_d=9$ MeV, $m_s=180$ MeV, $m_c=$
1.4 GeV, $m_b=$ 6 GeV, understood to be defined at a
renormalization scale of $1$ GeV.
%6
\item
A. Buras and M. Harlander, Max Planck Institute preprint
MPI-PAE/PTH, January 1992, Munich.
%7
\item
J. Polchinski and M.B. Wise, Phys. Lett. 125B 393 (1983).
%8
\item
Y. Nir and U. Sarid, Weizmann preprint WIS-92/52/Jun-PH
(1992).
%9
\item
L. Wolfenstein, Phys. Lett 164B 170 (1985); J.F. Donoghue et
al., Phys. Rev.
D33 179 (1986).
%10
\item
H. Georgi, Harvard University preprint HUTP-92/A049 (Aug.
1992).
%11
\item
T. Ohl, G. Ricciardi and E. Simmons, Harvard University preprint
HUTP-92/A053 (Dec. 1992).
%12
\item
The decay  $t\rightarrow c h^0$ has also been considered recently by
W.-S. Hou, Phys. Lett. B296 179 (1992), where relaxing the symmetry of
reference 1 is motivated by models with a Fritzsch-like texture.

%13
\item
The large value of $\epsilon_{U_2}$ is due to the fact that
$m_c/m_t$ is not
very much less than $V_{cb}$.

\end{enumerate}
\end{document}